\documentclass[12, preprint]{aastex}
\usepackage{natbib}
\usepackage{subfigure}

\newcommand{\captionfonts}{\small}

\makeatletter  
\long\def\@makecaption#1#2{%
  \vskip\abovecaptionskip
  \sbox\@tempboxa{{\captionfonts #1: #2}}%
  \ifdim \wd\@tempboxa >\hsize
    {\captionfonts #1: #2\par}
  \else
    \hbox to\hsize{\hfil\box\@tempboxa\hfil}%
  \fi
  \vskip\belowcaptionskip}
\makeatother   

\newcommand{\noprint}[1]{}

\begin{document}

\title{Rapidly accreting white dwarfs as supernova type Ia progenitors} 
\author{Kelly Lepo \& Marten van Kerkwijk} 
\affil{University of Toronto} 
\email{lepo@astro.utoronto.ca, mhvk@astro.utoronto.ca}

\begin{abstract} 
The nature of the progenitors of type Ia supernovae is still a mystery. While plausible candidates are known for both the single degenerate and double degenerate models, the observed numbers fall significantly short of what is required to reproduce the type Ia supernovae rate.

Some of the most promising single-degenerate type Ia progenitors are recurrent novae and super-soft sources (SSS). White dwarfs with higher mass transfer rates can also be type Ia supernova progenitors. For these rapidly accreting white dwarfs (RAWD), more material than is needed for steady burning accretes on the white dwarf, and extends the white dwarf's photosphere. Unlike super-soft sources, such objects will likely not be detectable at soft X-ray energies, but will be bright at longer wavelengths, such as the far ultraviolet (UV). Possible examples include LMC N66 and the V Sagittae stars.

We present a survey using multi-object spectrographs looking for RAWD in the central core of the SMC, from objects selected to be bright in the far UV and with blue far UV-V colors. While we find some unusual objects, and recover known planetary nebula and WR stars, we detect no candidate RAWD. The upper limits from this non-detection depending on our expectations of what a RAWD should look like, as well assumptions about the internal extinction of the SMC. Assuming they resemble LMC N66 or fainter versions of Wolf-Rayet stars we set an upper limit of 10 --14 RAWD in the SMC. However, our survey is unlikely to detect objects like V Sge, and hence we cannot set meaningful upper limits if RAWD generally resemble these.
\end{abstract} 


\maketitle

\section{Introduction}
Type Ia supernovae (SN Ia) serve as standard candles in extragalatic astronomy. They are thermonuclear explosions in carbon-oxygen white dwarfs that are generally thought to be triggered by the white dwarf approaching or exceeding the Chandrasekhar limit. A white dwarf can reach this limit by mass transfer from a companion, but the nature of the typical companion remains to be determined. Generally, two scenarios are considered. In the first, the ``single degenerate'' scenario, the companion is a normal, non-degenerate star, and mass is transferred via Roche-lobe overflow (or a strong wind). In the second, the ``double degenerate'' scenario, the companion is another white dwarf, and unstable mass transfer between and a subsequent merger of the two white dwarfs leads to an explosion.

For both scenarios, the observed number of suitable progenitor systems is significantly smaller than what is required to match the SN Ia rate. In particular, for the DDs, the SPY survey has failed to find the rather large expected number of binaries with a combined mass above the Chandrasekhar mass \citep{Napiwotzki07}, and hence the DD scenario may only be viable if systems with a lower combined mass may be made to explode \citep{Van-Kerkwijk10,Badenes12}.

For all single-degenerate progenitors, there must be enough mass in the binary system to put the white dwarf over the Chandrasekhar limit. In addition, a sufficiently large fraction of this mass should not just reach the white dwarf (rather than be lost in winds), but also be processed and retained (rather than lost in outbursts). Furthermore, the system must also transfer mass on a short enough time scale to match the observed SN Ia delay time distribution.  Finally, no strong mass loss should occur for the majority of progenitor systems, since neither most type Ia supernova nor most of their remnants show signs of interaction with dense circumstellar medium \citep{Panagia06,Chomiuk12,Badenes07}.

As outlined in \citet{Nomoto07}, depending on the accretion rate $\dot{M}$, nuclear burning on white dwarfs occurs in three modes. In most known systems, cataclysmic variables and symbiotics, white dwarfs accrete below the steady nuclear burning rate ($3.066 \times 10^{-7} (\frac{M}{M_{\odot}}-0.5357) M_{\odot} {\rm\ yr^{-1}}$), which produces novae. The slow accretion rate results in a shell that, at the time hydrogen burning ignites, is degenerate and thin, and hence cannot expand sufficiently to counteract the heating due to fusion, leading to thermonuclear runaway \citep{Warner95, Prialnik86}.

It is unknown whether the white dwarf gains or loses mass over time as a nova system undergoes cycles of periodic, unstable nuclear burning, but present evidence suggests it is difficult to grow the white dwarf. From numerical simulations, \citet{Yaron05} find that only nova systems with high accretion rates ($\sim 10^{-8}\ldots10^{-7} M_{\odot}{\rm\ yr^{-1}}$) will retain mass. Recurrent novae are perhaps the best candidates for type Ia progenitors because they have a relatively large $M_{\rm WD}$ and high  $\dot{M}$. But they are rare, with only 10 known in our Galaxy \citep{Schaefer10}.  

The most popular single-degenerate progenitors are systems where accretion rates are within a narrow range, of about a factor of 2.5 in $\dot{M}$, where stable nuclear burning occurs on the surface of the white dwarf (between the limit listed above an $6.68\times 10^{-7} (\frac{M}{M_{\odot}}-0.4453)M_{\odot} {\rm\ yr^{-1}}$). Unlike in a nova, the hotter hydrogen shell is thicker and sufficiently non-degenerate to allow stable burning. For white-dwarf masses above $\sim\!0.8\,M_\odot$, the temperatures become high enough for the peak of the spectrum to move to the soft X-ray range, hence these systems are referred to as super-soft sources \citep[SSS,][]{van-den-Heuvel92}. With fast accretion and no mass loss, white dwarfs in such systems can grow to the Chandrasekhar mass within an interval that reproduces the known delay time distribution, as long as the initial WD is sufficiently massive \citep{Di-Stefano10}.

Unfortunately, the number of SSS observed is too small: both total X-ray flux measurements in nearby elliptical galaxies \citep{Gilfanov10} and counts of SSS in nearby  galaxies \citep{Di-Stefano10} find that the number of SSS are a factor of 10 to 50 short of what is needed to reproduce the SN Ia rate.

Thus, either single degenerates do not produce (most) SNe Ia, explosions are triggered at sub-Chandrasekhar masses, or single degenerates are in the SSS phase only for a short time \citep{Woosley94}. According to \citet{Hachisu10}, the latter is expected: they argue the X-ray fluxes observed in elliptical galaxies are consistent with those expected for the single degenerate scenario, as long as one takes into account that the SN Ia progenitor will be a SSS only part of the time, after a phase in which the accretion rate is above that required for producing SSS, and followed by phase as a recurrent nova.  In the context of this scenario, the absence of evidence for dense circumstellar medium may also not be surprising.

The above recognizes the possibility that white dwarfs can accrete at a rate above the steady nuclear burning rate. Such white dwarfs will have a stable hydrogen burning layer over the degenerate core, with an extended envelope. The WD should retain all of the material it fuses on its surface \citep[barring any eruptions caused by instabilities in the He layer, see][]{Idan12} and the high accretion rate will give such systems a sufficiently prompt delay time. They will not be super-soft X-ray emitters, since their envelope and/or winds cause the photosphere to move out, and the emission to peak at longer wavelengths, likely the (far) ultraviolet (UV).  For lack of a better name, we dub these systems rapidly accreting white dwarfs (RAWD).

There are two general ideas for what a RAWD might look like. The simplest and perhaps naive one would be that the white dwarf tries to become a red giant, but is limited by the orbit, and loses mass from the L2 Lagrange point and/or forms a common envelope \citep{Nomoto79,Iben88}. Given that most cataclysmic variables have periods of about 80 min to 10 hours, one thus expects the envelope to be at most a few solar radii in size. With an expected luminosity of $\sim\!10^{4.5}\,L_\odot$ \citep{Nomoto07}, the effective temperature would be $\sim\!10^{4.5}$ to $10^5\,$K, i.e., the sources might look like undersized OB stars. If the accretion rate is high enough for the white dwarf overflow to its Roche lobe, the system will form a common envelope. While such systems may be interesting on their own, they will not be type Ia progenitors. 

The second possibility is that the white dwarf will lose mass through optically thick winds, produced by the same strong peak in opacity (largely due to iron lines) at $T\simeq150,000$ K that drives winds in late-time nova envelopes \citep{Hachisu96}. Depending on the mass loss rate and the accretion rate, the white dwarf may or may not fill its Roche lobe. With such a wind, these sources may appear like Wolf-Rayet stars or [WR] planetary nebula (of WN subtype, i.e., showing the products of the CNO cycle).

In either case, one might hope to identify a population of rapidly accreting white dwarfs by looking for sources that are bright in the ultraviolet and are hot, but are fainter than main-sequence stars of similar temperature. Here, after describing sources suggested to be RAWD, we present our search for such UV-bright sources in the Small Magellanic Cloud.

\section{Objects suggested to be Rapidly Accreting White Dwarfs}\label{sec:rawdcandidates}
Among the objects suggested to be rapidly accreting white dwarfs (sometimes referred to as ``accretion wind evolution"  systems), the ones studied in most detail are the V Sge stars and the [WN] planetary nebula LMC N66. 

\subsection{V Sge stars}
V Sge is a peculiar double-lined binary that shows clear eclipses in its light curve, with a period of 0.541 days. There is no consensus on the exact nature of the system, and several models have been proposed to explain its light curve. Most relevant for our purposes is the suggestion by \citet{Hachisu03a} that V Sge may be a white dwarf accreting at a high rate from a Roche-lobe filling companion. Their models  assume a 1.25 $M_{\odot}$ WD and a 3.0-3.5 $M_{\odot}$ companion, which would make the system a excellent type Ia progenitor. 

V Sge shows quasi-periodic optical high and low states of $\sim 180$ days, and is a SSS in the low state. In the optical high state, the WD accretes faster that it can burn hydrogen, and thus is a RAWD.  \citet{Hachisu03a} suggest that during this state, the accretion disk around the white dwarf grows, becoming circumbinary and absorbing soft X-rays produced by the white dwarf. The white dwarf also accumulates matter on its surface, creating an extended envelope. The expanding photosphere pushes the peak of the white dwarf's emission to longer wavelengths.  During the optically low state, winds from the white dwarf suppress the mass accretion rate from the companion. The disk shrinks, as does the envelope of the white dwarf. This brings the peak of the emission from the white dwarf to shorter wavelengths, and it should behave much like a SSS. The smaller, less flared disk is less luminous, and thus contributes less to the total luminosity of the system. 

The spectra of V Sge show numerous strong emission lines of highly ionized species, including both  \ion{O}{6} (3434, 3811, 5290, 5584 \AA) and \ion{N}{5} (4945, 4603 \AA). These lines are not present in  most of the known Galactic W-R stars. It also has a \ion{He}{2} 4686 line that is about twice as strong as its H $\beta$ line, indicating high degree of ionization.  No \ion{He}{1} is seen, which suggests the lines are formed by photoionization \citep{Steiner98}.

Stars with similar spectra, including QU Carinae, WX Cen, V617 Sgr, and RX J0513.9-6951, are considered members of the ``V Sagittae'' class of nova-like cataclysmic variables \citep{Steiner98}. While not as well studied, it is assumed that these objects are physically similar to V Sge. In particular, \citet{Hachisu03} model the transient super-soft X-ray source RX J0513.9-6951 using the same disk/wind model used for V Sge, and attribute differences in their observed properties to different viewing geometry. 

\subsection{LMC N66}
LMC N66 (also known as WS 35 and SMP 83) is a unusual planetary nebula with an early [WN] core, located in the Large Magellanic Cloud.  It undergoes periodic outbursts, where its luminosity increases by about an order of magnitude for a period of years. In its quiescent state, its luminosity ($\log[L/L_{\odot}]= 4.5$) exceeds that of known [WC] planetary nebulae, and in outburst its luminosity ($\log[L/L_{\odot}]= 5.4$) is on par with galactic WN stars \citep{Hamann03}. Two outbursts have been observed, with peak luminosities occurring in 1994 and 2007 \citep{Pena95,Hamann03,Pena08}.  There is no evidence in archival data that any outbursts occurred from 1955 to 1990, and it is not detected as an X-ray source \citep{Hamann03}.  We conclude that it is in its high state at most half the time, and never reaches a state similar to a SSS.

Spectra of LMC N66 show incompletely CNO-processed material --- strong helium emission lines, some hydrogen, enhanced nitrogen and depleted carbon. From the width of the lines one infers a terminal wind speed $v_{\infty} = 2200{\rm\,km\,s^{-1}}$ during outburst and  $1600{\rm\,km\,s^{-1}}$ during quiescence. The model that has the fewest contradictions with the observed properties of LMC N66, is that the central object of the planetary nebula is a low-mass binary, with a WD primary and a non-degenerate secondary. In this model, the WD rapidly accretes mass, and loses some to winds, in other words, it is a RAWD. The outbursts --- hard to explain with a single star model --- may be due to helium shell flashes on the white dwarf \citep{Hamann03}. 

Very little work has been done to verify if the core of LMC N66 is in fact a binary. \citet{Hamann03} find that there is no radial velocity variations in the observed spectra that would indicate a binary system. However, given the widths of the emission lines and the confusion by nebular material surrounding the central objects(s), this may not be surprising. Unless there is a systematic search for RV variations, looking preferentially at lines that originate from the irradiated atmospheres of a main-sequence companion, it is unlikely that one will see evidence of a periodic signal (see, e.g,. \citet{Miszalski11} for a successful spectroscopic detection of a binary-core planetary nebula). 

Assuming the system consists of a white dwarf and a normal star, models by \citet{Hamann03} put the radius of the photosphere of the WD at $0.52-1.38\,R_{\odot}$. This is about what one would expect for the Roche lobe of a white dwarf in a close binary. They find the mass loss rate of the object varies from $\dot{M} = 10^{-5.7}$ during quiescence to $\dot{M} =10^{-5.0}\,M_{\odot}{\rm\,yr^{-1}}$ during outburst. This is would put the white dwarf in the high $\dot{M}$ regime. Both the radius and $\dot{M}$ estimates are consistent with our ideas of RAWD.

While V Sge seems to only be a RAWD during its optically high state, the cooler surface temperature and lack of X-ray detection of LMC N66 suggests it is a RAWD full time. This may be because the average mass loss rate from the secondary, and hence the accretion rate onto the WD and the mass loss rate from winds, is higher in LMC N66 than in V Sge. 

\section{Expected numbers of RAWD \label{sec:numbRAWD}}

In order to search for RAWD, it is helpful to have dense but still resolvable stellar fields, so that many stars can be scrutinized, low extinction along the line of sight, allowing the ultraviolet to pass through, and known distance, so that one can use both color and luminosity for selection.  The Magellanic Clouds fulfill these criteria best, and we picked the Small Magellanic Cloud (SMC) as it had better UV coverage.

To estimate how many RAWD might exist if single degenerates dominate the SN Ia rate, we follow \citet{Di-Stefano10} and first calculate the number of accreting WDs expected in the SMC, based on a SN Ia rate estimated from the galaxy's blue luminosity (Eq.~4 from her work),
\begin{eqnarray}
N_{\rm acc} = 1500\left(\frac{\Delta M}{0.4M_{\odot}}\right) 
      \left(\frac{8 \times10^{-7 }M_{\odot}{yr}^{-1}}{\beta  \dot{M}_{\rm in}}\right) 
      \left(\frac{L_B}{10^{10} L_{\odot}}\right),
\end{eqnarray}
where $\dot{M}_{\rm in}$ is the mass transfer rate and $\beta$ is a retention factor, which is a function of $M_{\rm WD}$, the mass of the white dwarf, and  $\dot{M}_{\rm in}$. The maximum accretion rate implicit in the derivation of this equation is $\beta\dot{M}_{in} = 8 \times 10^{-7}\,M_\odot{\rm\,yr^{-1}}$ for a solar mass white dwarf (or, equivalently, a 0.4  $M_{\odot}$ gain in mass and an accretion lifetime of $5\times 10^5$ years). This is approximately the upper limit for steady hydrogen burning on the white dwarf and thus the most rapid rate that mass can be added to the white dwarf (for higher accretion rates, the excess is lost through winds or outflows). Stars with smaller accretion rates than the steady burning limit will have longer lifetimes, and thus there will be more accreting systems to observe. Setting $\beta\dot{M}_{\rm in} = 8 \times 10^{-7} M_\odot{\rm\,yr^{-1}}$ gives a rough lower limit on the number of accreting white dwarfs that we should observe.  Using $\Delta M=0.2$ to $0.8\,M_\odot$, corresponding to initial masses from $1.2\,M_\odot$ (the maximum possible for a CO white dwarf) to $0.6\,M_\odot$ (the typical mass of WDs formed formed in isolation), and $L_B=4.4\times10^8\,L_\odot$ for the SMC, we find that the number of accreting white dwarfs needed to reproduce the SN Ia rate is $N_{\rm acc}=33\ldots132$.  

Since there are only four (possibly five) known SSS in the SMC \citep{Greiner00,Sturm11}, there should be dozens of missing progenitors in the SMC if single degenerates produce most SN Ia.  From the models of \citet{Hachisu08,Hachisu10}, such progenitors spend only 10\% of their time as SSS; for all systems that have a secondary with a mass greater that $\sim\!2.0\,M_{\odot}$, most of the accretion will happen in a wind (or rapid accretion) phase early in their evolution that lasts $\sim 5 \times 10^{5}\,$yr, i.e., while the sources are RAWD, before the accretion rate drops and the object becomes a SSS and later a recurrent nova. 

Of course, even if single degenerates are not responsible for most SN Ia, one would expect some rapidly accreting white dwarfs with high accretion rates to exist, in numbers likely not too dissimilar from those of the SSS.


\section{A hunt for RAWD in the SMC}
While some examples of possible RAWD are known, we cannot be sure these sample the full array of possibilities.  Generally, however, one expects the emission to peak in the UV, since the luminosity is high, yet the effective radius relatively small, as it is unlikely to exceed the Roche lobe.  Hence, we aimed to select candidates that had unusual UV emission, indicative of a hot component.  (We will return to estimates of the effectiveness of our estimates in Section~\ref{sec:expectations}.)

The Magellanic clouds were surveyed by the Ultraviolet Imaging Telescope (UIT) during two space shuttle missions in late 1990 and early 1995, using several FUV filters, including B5 at 162 nm \citep{Cornett97}.  Since the UIT observations cover the entire central bar of the SMC, while the LMC observations are in scattered fields, we chose to use the SMC data to simplify followup observations.\footnote{The central regions of both Magellanic clouds are too UV bright for {\em GALEX}.}

\subsection{Selecting candidate sources}
We matched objects in the UIT SMC catalog with the optical Magellanic Cloud Photometric Survey \citep[MCPS,][]{Zaritsky02}. Since the astrometry given in the two catalogs is not on exactly the same system, we redid the astrometry using an iterative procedure, in which we: (1) found the object with the brightest U magnitude in the MCPS catalog within a given radius of each UIT point; (2) scaled, rotated, and translated all of the points in the UIT field, performing a least squared fit to minimize the distance between the matched UIT and MCPS points; and (3) decreased the search radius and repeated steps (1) and (2).

We first ran this algorithm on all points in the MCPS catalog with $U<16$, and beginning with the resulting, revised positions, ran it on  all points in the MCPS catalog with $U<19$. We began with a search radius of 20 arcsec and finished with a radius of 3 arcsec.  Using these final matches between the catalogs, we took the position of each point to be the position of the brightest source in MCPS catalog within 3 arcseconds of a UIT point. 

Unfortunately, due to a programming error, we actually selected the faintest rather than the brightest star in step (1), which effected our astrometric solution as well as the matches between the two catalogs. We chose our fiber positions before we noticed this error. However, since there are not that many MCPS points with $U<19$ around each UIT point, this does not seem to have had a large effect. Re-running the matching code to correctly select the brightest star, we see almost identical numbers of matches between the two catalogs (with 6230 matches originally, 6228 with the correction). The same MCPS point was selected around a UTI point for  5907 sources, meaning a different point was selected only for about 5 percent of all points in our catalog. Our online data includes the revised UIT/MCPS catalog, and includes all MCPS points within 3 arcsec of the UIT point. From here on, we will only consider those source for which the identification did not change.


\begin{deluxetable}{llcccccclllllllllllllllll}
\rotate 
\tabletypesize{\scriptsize}
\tablewidth{1.45\linewidth}
\setlength{\tabcolsep}{0.032in} 
\tablecaption{Combined UIT and MCPS catalog \label{SMC_table} }
\tablehead{ 
\colhead{USS cat} &
\colhead{UIT} &
 \colhead{UIT} &
 \colhead{rev. UIT} &
 \colhead{rev. UIT} &
\colhead{$m_{162}$}&
\colhead{err}&
 \colhead{UIT} &
 \colhead{MCPS}&
 \colhead{MCPS} &
 \colhead{mU} &
 \colhead{ err} &
 \colhead{mB} &
 \colhead{ err} &
  \colhead{mV} &
 \colhead{ err}&
  \colhead{mI} &
 \colhead{ errI} &
  \colhead{mJ} &
 \colhead{ err} &
  \colhead{mH} &
 \colhead{ err}&
 \colhead{mK} &
  \colhead{ err} &
  \colhead{r} \\
   \colhead{name} &
 \colhead{ra (d)} &
 \colhead{dec (d)} &
 \colhead{ra (d)} &
 \colhead{dec (d)} &
\colhead{}&
\colhead{}&
 \colhead{field} &
 \colhead{ra (d)}&
 \colhead{dec (d)} &
  \colhead{} &
 \colhead{} &
 \colhead{} &
 \colhead{} &
  \colhead{} &
 \colhead{} &
  \colhead{} &
 \colhead{} &
  \colhead{} &
 \colhead{} &
\colhead{} &
 \colhead{}&
 \colhead{} &
  \colhead{} &
  \colhead{(asec)} 
}
\startdata
 &11.4488 &-72.9392 &11.4517 &-72.9381 &14.76 &0.08 &f1&11.4511 &-72.9376 &16.46 &0.04 &17.19 &0.03 &17.23 &0.04 &17.39 &0.08 &  &  &  &  &  &  &2.02\\
 &12.1096 &-73.2508 &12.1053 &-73.2508 &15.26 &0.12 &f1&12.105 &-73.2506 &16.68 &0.04 &17.31 &0.03 &17.36 &0.06 &17.34 &0.04 &  &  &  &  &  &  &0.95\\
 &12.32 &-73.2517 &12.316 &-73.2517 &13.53 &0.05 &f1&12.3153 &-73.2521 &17.18 &0.13 &17.53 &0.03 &17.0 &0.07 &18.13 &0.12 &  &  & & & & &1.5\\
&12.1133&-73.0264&12.1149&-73.0258&14.12&0.07&f1&12.1144&-73.0262&16.07&0.04&16.91&0.02&17.04&0.03&17.15& 4& & & & & & &1.48\\
&12.2925&-73.2525&12.2884&-73.2526&14.08&0.05&f1&12.2887&-73.2527&15.64&0.03&16.43&0.02&16.46&0.03&16.69&0.04& & & & & & &0.47\\
Bkrnd 15 121&12.7192&-73.2392&12.716&-73.2394&12.61&0.09&f1&12.7162&-73.2395&13.78&0.03&14.65&0.02&14.66&0.04&14.84&0.08&15.01&0.05&15.0&0.08&15.12&0.15&0.31\\
Bkrnd GT15 492&11.7038&-73.1047&11.7028&-73.1042&13.46&0.12&f1&11.7035&-73.1039&15.03&0.03&15.88&0.02&15.95&0.03&16.14&0.04& & & & & & &1.31\\
\enddata 
\tablecomments{Table 1 is published in its entirety in the electronic edition of the Journal. A portion is shown here for guidance regarding its form and content.}
\end{deluxetable}

\begin{figure}
\begin{center}
 \includegraphics[width= 1\linewidth]{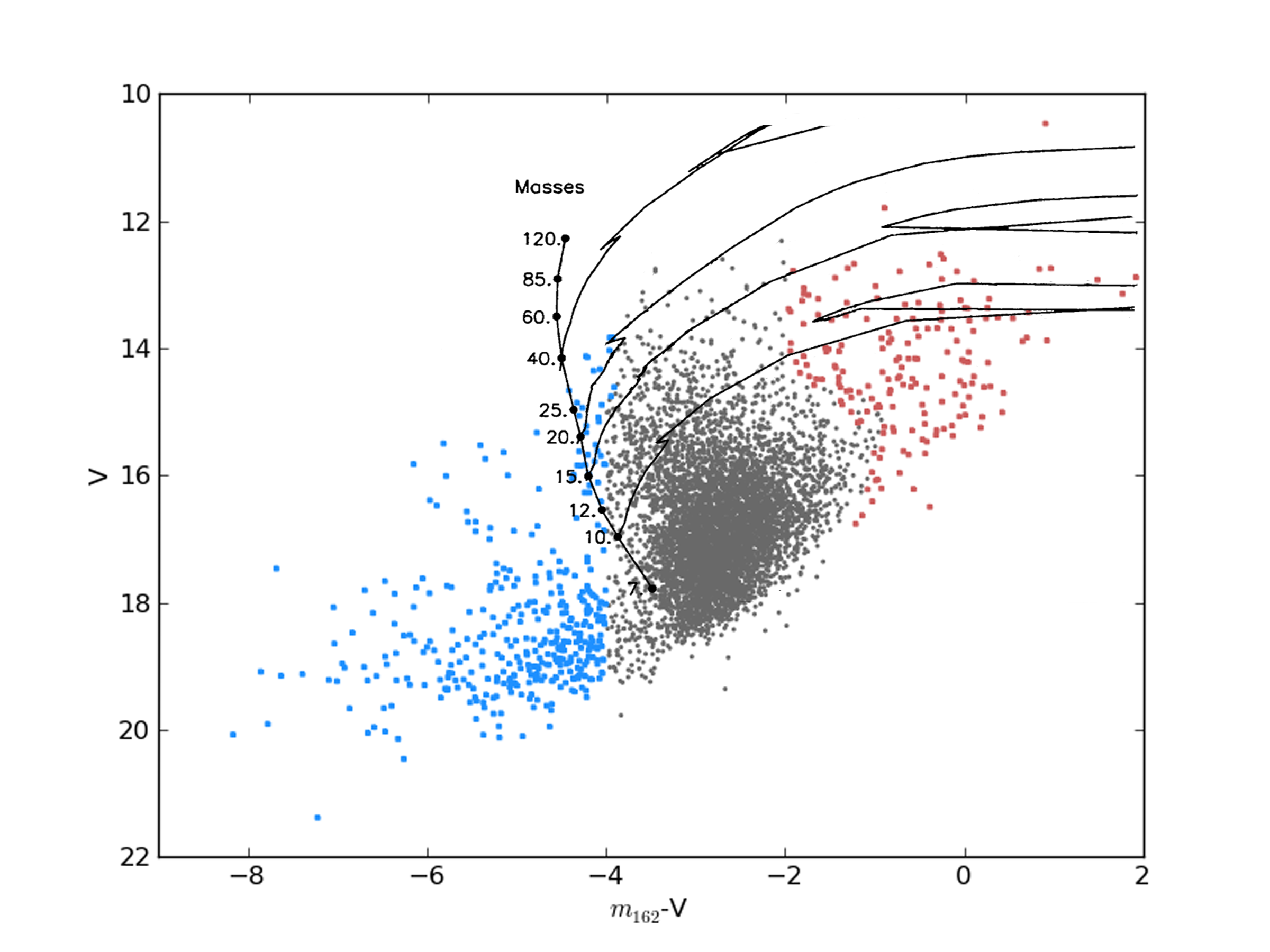}
 \caption{Ultraviolet to visual color-magnitude diagram for stars in the bar of the SMC, using values from UIT and MCPS. There are 5670 sources in the central clump (circles), 175 sources redder than the central clump (squares) and 385 sources bluer than the central clump (squares).  Overlaid is are evolutionary tracks of OB stars (Fig.~4 from \citealt{Cornett97}).}
   \label{cmd}
\end{center}
\end{figure}

The color-magnitude cuts for candidate sources were chosen to maximize the number of objects with a possible UV excess, while minimizing the number of normal main sequence stars in our survey. Figure~\ref{cmd} is a color magnitude diagram of $m_{162}-V$ vs $V$ (here, magnitudes are on the ST system). Overlaid on this plot is the position of the main sequence calculated by \citet{Cornett97}. 

The central, main sequence clump is by definition the densest region of Fig. \ref{cmd}. Thus, rather than using a color cut to identify the outliers, we used a density cut. To do this we calculated the number of neighboring points around each point in Fig. \ref{cmd}, within a radius of 0.3 mag in color-magnitude space.  We selected candidate points that were more than $1.35\sigma$ outliers in the number of neighboring points. This essentially amounted to a cut in color of $m_{162}-V<-4$ for the points with a UV excess (shown in blue in Fig.~\ref{cmd}). For sources with a UV deficit, however, the color limit varies much more with brightness (shown in red).

\begin{figure}
\begin{center}
 \includegraphics[width= 1\linewidth]{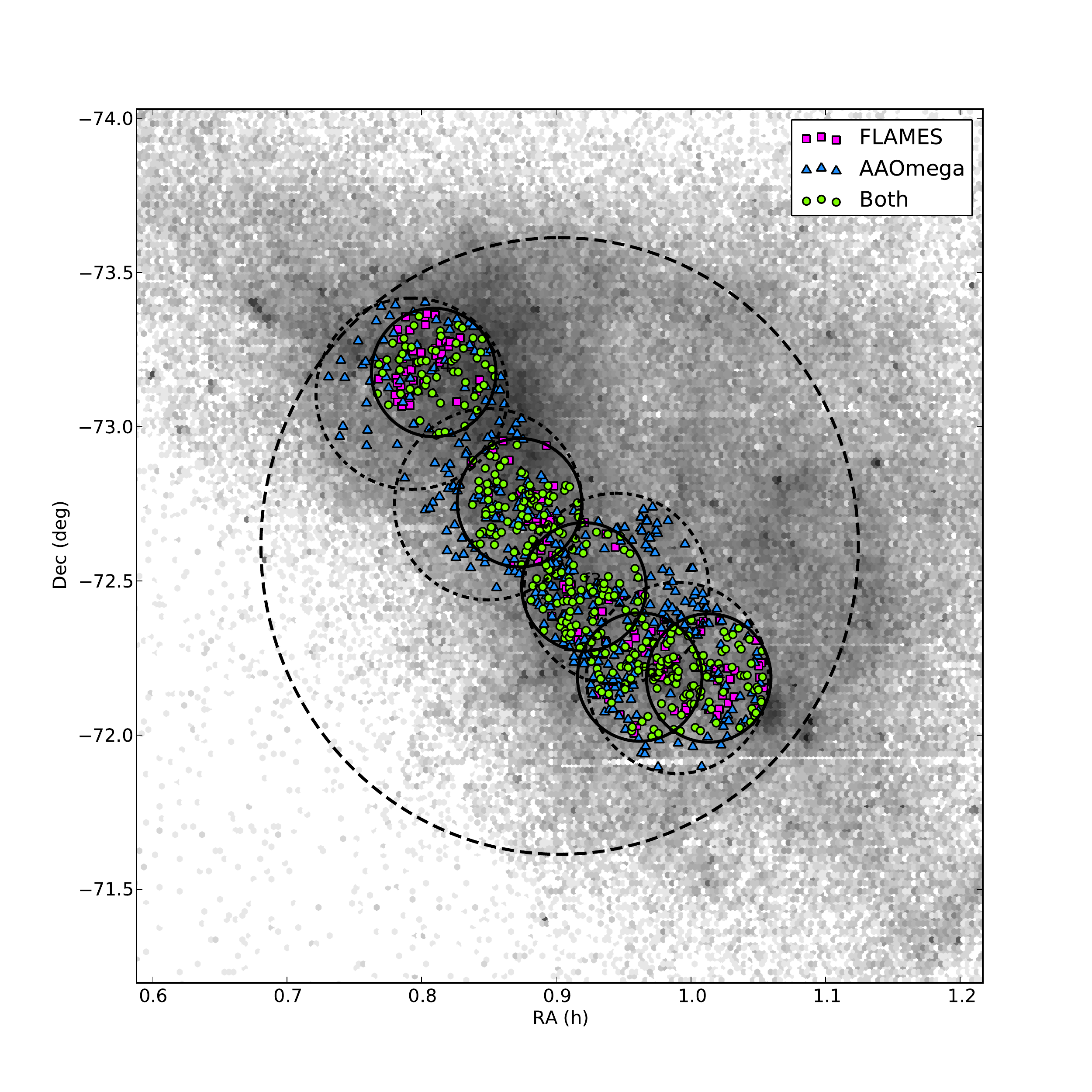}
 \caption{Objects observed in this survey.  In the background is a density map of objects in the SMC MCPS survey, with $V < 19$, which traces the stellar density of the SMC.  The dot-dashed lines represent the fields observed by UIT. The large, black dashed line is the AAOmega field, the smaller solid lines are the FLAMES fields. The purple square points are objects observe with FLAMES, the blue triangles are objects observed by AAOmega, and the green circular points are observed by both.}
   \label{circles}
\end{center}
\end{figure}

\subsection{Possible Mis-matches between the UIT and MCPS Catalogs}

An inherent problem in trying to find outliers by matching two catalogs of astronomical objects taken at two different wavelengths at two different times is that one will find not just objects with a true color excess but also variable sources and bad matches.  For instance, for some of the objects that were outliers based on their $m_{162}-V$ color, we find unremarkable spectra, and it seems likely the UV emission was instead from a much brighter object just beyond our selection radius (for which typically the color would place it in the central, main sequence clump in Fig.~\ref{cmd}).  

To get a handle on the influence of mis-matches, we looked at each object in the MCPS (optical) catalog we matched to each UIT (FUV) point, as well as at the second closest star in the MCPS catalog. The second closes star should, in general, be the most likely match if the initial one was incorrect. 

For our initial match, 233 of 5907 objects (4\%) were bluer than the central clump of the CMD. Using the second closest MCPS point, we get 3333 out of 5907 objects (56.4\%) bluer than the central clump. Clearly, the first match was much better. Using the second closest MCPS point moves 157 (67.4\%) of the objects that were originally ``blue'' candidates into the central clump. 

For comparison, we also assigned each UIT point a random point from the MCPS catalog. If the matching program were perfect, the second closest point should be identical to a random point, as there would be no miss-matches and hence no special significance to the second closest point in the MCPS catalog. After 500 runs, using the V magnitude from a random MCPS point and the original UIT m$_{162}$ magnitude, 3251 ($\pm34$) of 5907 objects (55\%) were bluer than the central clump. Assigning a random point produces an almost identical CMD to the one produced using the second closes point, within errors. It seems that our matching program did a good job for the majority of the sources in the UIT catalog. However, using a random MCPS point only moves 63 ($\pm 6$, 2\%) of the objects that were originally bluer than the central clump into the central clump. The improvement seen using the second closest point over a random point suggests that  $\sim\!40\%$ of the original bluer outliers were due to bad matches between the UIT and the MCPS catalogs. Conversely, this suggests that $\sim\!60\%$ of the blue outliers should be true matches.  Of course, this does not prove a true excess (instead of measurement errors or variability), but does suggest that any RAWD with properties consistent with our selection criteria would likely be selected. 

\section{Observations}
We took optical spectra of a majority of the objects outside of the main-sequence clump of Figure~\ref{cmd}, using the multi-object spectrometers FLAMES on the VLT and AAOmega on the Anglo-Australian Telescope (AAT) \citep{Pasquini02,Sharp06}. We also took spectra of objects within the central clump using any extra fibers.

With FLAMES, we observed five, $25\arcmin$ diameter fields, which overlapped slightly, with $\sim\!110$ objects in each field. Two fields were observed twice because the first observations were not taken within specified seeing conditions, but even these observations turned out to be usable. In the end, we had 503 unique objects and 745 spectra taken from Oct 2010 to Jan 2011.

The spectra were taken using Giraffe with its LR03 setting, covering 4501--5078\,\AA\ at a resolving power of $\lambda/\Delta \lambda \approx 7500$. This setting covers H$\beta$, several \ion{He}{1} and \ion{He}{2} lines (including \ion{He}{2} at 4686\,\AA), [\ion{O}{3}] at 4959 and 5007\,\AA, as well as the Bowen blend of \ion{N}{3} and \ion{C}{3} lines at 4640\,\AA. We chose this band as a compromise between being able to do stellar classifications of hot stars (H$\beta$, \ion{He}{1} and/or \ion{He}{2} absorption lines), and having features relevant for ionized gas, be it nebular (strong, narrow H$\beta$ emission lines with strong [\ion{O}{3}] lines) or from a wind powered by a hot central core (like WR stars or RAWD; strong \ion{He}{2}, N and/or C lines).

The two degree diameter field of view of AAOmega allowed us to image the entire core of the SMC in one field. We took three observations of the same field, but targeting objects in different magnitude ranges  with different exposure times. This left us with 857 spectra of 732 unique objects. Observations were taken from 4-5 June 2011, covering 3700--5800\,\AA\ in the blue with the 580V grating, and 5600--8800\,\AA\ in the red with the 385R gratings.  The resolving power in each band is $\sim\!1300$.  The blue spectra include the range covered by the FLAMES observations, and we looked for similar spectral features.  In the red spectra, the most prominent line is H$\alpha$, which was useful for interpreting otherwise ambiguous spectra.

We assigned all objects in our combined UIT/MCPS catalog a priority based on their positions in the Figure~\ref{cmd} as well as their $B-V$ colors. In order of  priority these were (1) stars bluer than the central clump with a $B-V < 0$, (2) stars bluer than the central clump with a $B-V > 0$, (3) stars redder than the central clump, and (4) all stars in the central clump. Stars were assigned fibers according to their priority and the instrumental limitations on fiber positions.

Of the 233 candidates bluer than the central clump , we observed 204 (88\%; 14 only with FLAMES, 71 only with AAomega, 119 with both). Of the 171 objects redder than the central clump, we observed 150 (88\%; 19 only with FLAMES, 52 only with AAomega, 79 with both). Finally, of the 5503 candidate objects from the central clump, we observed 376 (6.8\%; 116 only with FLAMES, 206 only with AAomega, 54 from both). Limitations on fiber positions prevented us from observing 100 percent of the blue candidate sources.  

\begin{figure}[h]
\begin{center}
 \includegraphics[width= 1.\linewidth, angle=90]{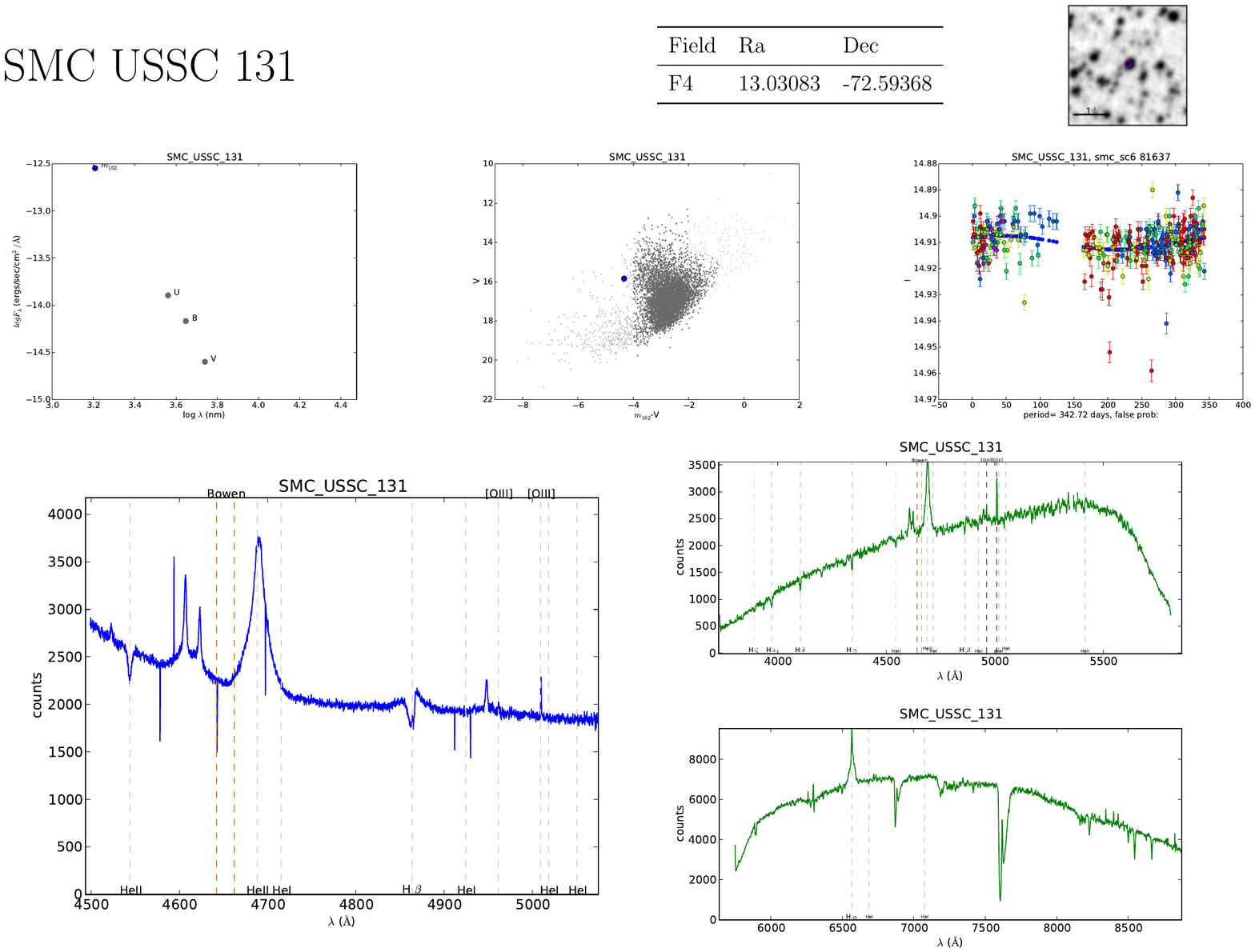}
 \caption{Summary of the observations for SMC USSC 131. {\bf Top row:} Name of object in the survey. Survey field, ra (degrees), dec (degrees). Digital sky survey image of the area around the UV source, revised position of the UV source is shown in blue, position of the optical counterpart is shown in red. Circles are 3\,arcsec in diameter.  {\bf Second row:} Spectral energy distribution of the object, showing the UV flux and several optical bands from MCPS and IR bands from 2MASS (if available). Color magnitude diagram similar to Figure~\ref{cmd} showing the position of the object in the large blue circle. If available, Ogle II lightcurve with a sine curve at the best-fit period overdrawn in blue. {\bf Third row:} Spectra from FLAMES (left) and AAOmega (right). Wavelength is in \AA. [{\em See the electronic edition of the Journal for Figs. 3.1--3.965}]}
   \label{summary}
\end{center}
\end{figure}

We assigned all objects a rough classification by eye, using summary graphs such as those shown in Figure~\ref{summary}. In addition, since many cataclysmic variables including V Sge stars show eclipses, ellipsoidal variation or periodic irradiation effects due to the hot primary irradiating the secondary star, we looked for periodic variations using the Ogle microlensing survey of the SMC \citep{Szymanski05,Udalski97}. We found the most likely photometric period of the objects from FFT periodograms created by using the the Lomb-Scargle method on Ogle II photometry data (of objects that were within the Ogle SMC fields). We were able to recover several known eclipsing binaries from the Ogle data, but no objects with short-period variations had unusual spectra, and no objects outside of the central clump showed any periodic variations.

We did recover some known objects that are similar to what we expect a RAWD to look like (discussed in more detail in Appendix~\ref{sec:interesting}): SMC-SMP 8, a [WC 8] planetary nebula, as well as SMC WR11 and WR 12, the two known Wolf-Rayet stars in the central core of the SMC. These objects have strong winds and a UV excess, which give us confidence that we would see a RAWD if it had Wolf-Rayet like features.

As expected, the majority of objects in the central ``main sequence'' clump were normal OB stars along with Be and B[e] stars.  Stars to the red side of the central clump (that had abnormally low FUV emission) were generally O, B and A giants. 

\begin{figure*}
\begin{center}$
\begin{array}{cc}
\includegraphics[width= .7 \linewidth]{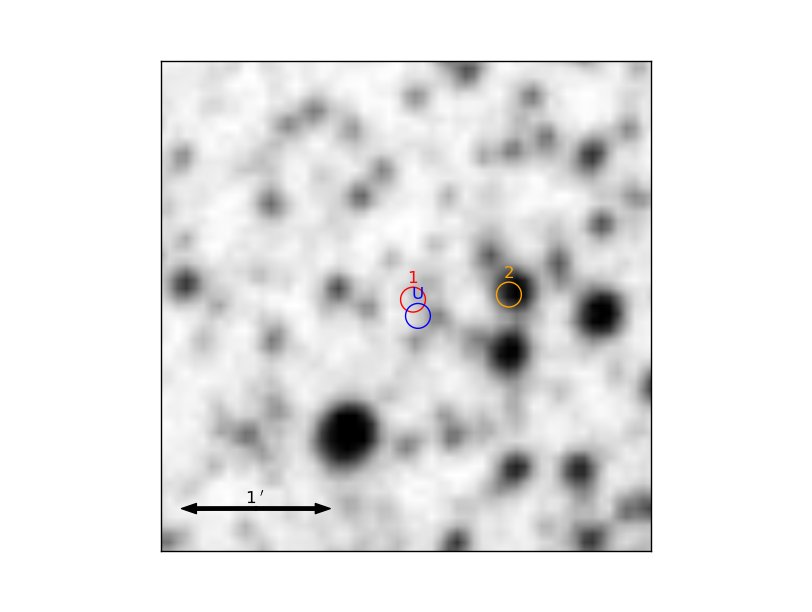}\\
\includegraphics[width= .6 \linewidth]{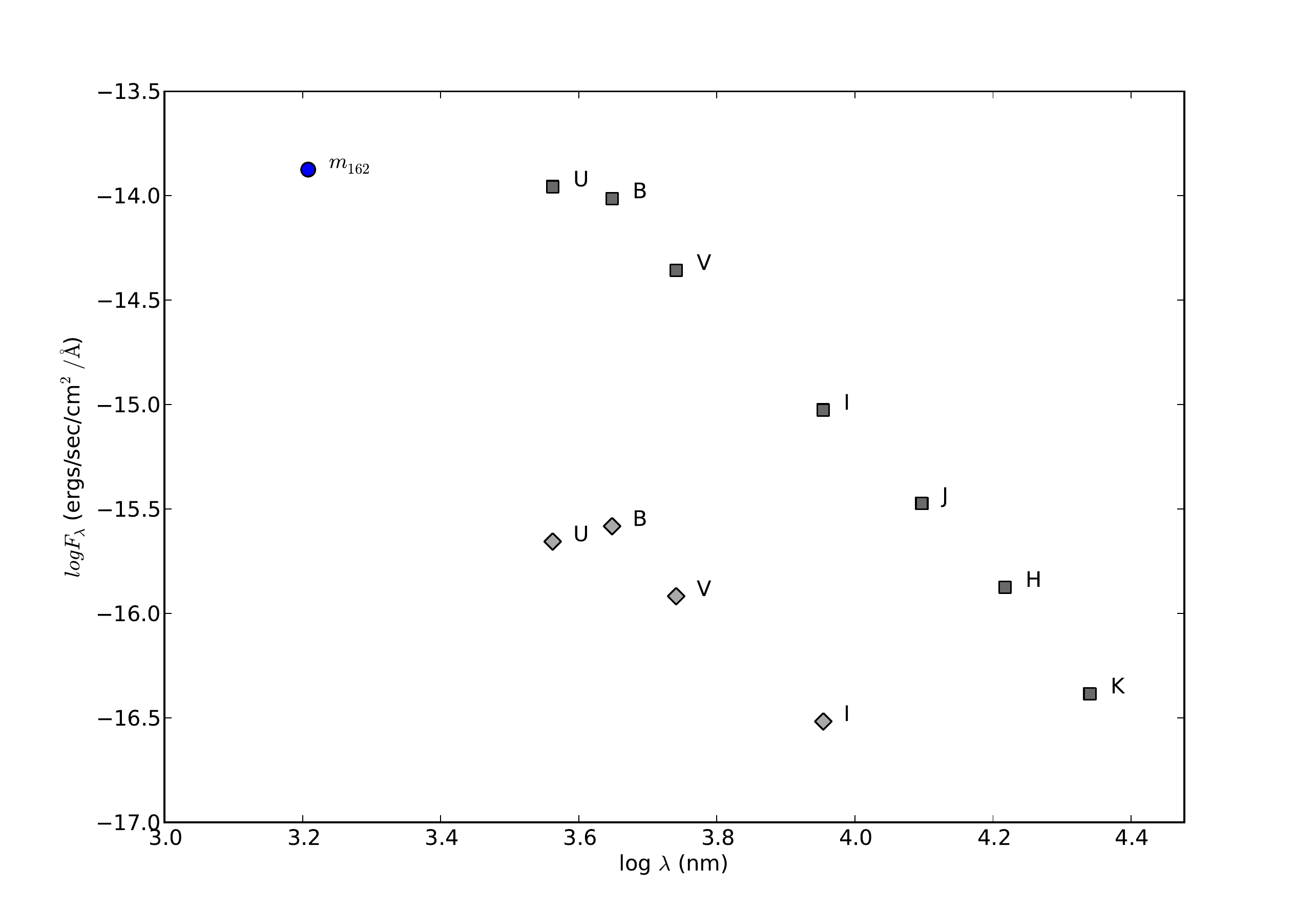}
\end{array}$
\end{center}
\caption{Example of a miss-match between the optical and UV catalogs. {\bf Top:} Image from the digitized sky survey. The position of the UIT UV source, using our revised astrometry, is shown with a blue 3\,arcsec radius circle labeled ``U". Our candidate optical counterpart, star 1, is shown with a red circle.  Most likely, the the ultraviolet source corresponds to one (or more) of the brighter stars, possibly the star marked with the orange circle, star 2. {\bf Bottom:} SEDs for star~1 (diamonds) and star~2 (squares),  with $U$, $B$, $V$, and $I$ from MCPS, $J$, $H$, and $K$ from 2MASS, and $m_{162}$ from UIT.  One infers a UV excess in the SED of star~1, but this is probably because the optical and UV fluxes come from different objects. Compared to all of the other nearby stars, the SED of star~2 is the most consistent with a single star that has the observed UIT $m_{162}$. Since the spectra of star~1 are unremarkable and inconsistent with a hot object, we conclude star~2 is the likely counterpart to the UV source.}
 \label{bad_match}
\end{figure*}

Stars to the blue side of the central clump (with a FUV excess) fell into a few broad categories. These include, in order of frequency, (A) About 50\% of the objects that show a UV excess in their SED are in a crowded field, yet have unremarkable spectra. They are most likely a mismatch between the FUV and optical catalogs.  An example is shown in Fig.~\ref{bad_match}. (B) About 20\% appear to be main sequence stars, some with nebular lines such H$\alpha$ and [\ion{O}{3}]  that may arise from the star or the surrounding ISM. These sources were almost exclusively found near the edge of the central clump, so they likely belong to that main sequence group.  (C) About 20\% appear to be misaligned fibers or objects with bad positions from the MCPS catalog --- objects that do not have a counterpart DSS images, have little flux and show spectra more consistent with the ISM rather those of stars.  And (D) the remaining 10\% which includes extended objects like star clusters and nebulae, objects with a bad $V$-band magnitude in the MCPS catalog (or at least one highly inconsistent with the object's $U$, $B$, and $I$-band magnitudes), as well as the three known WR stars discussed above.

Other than the two WR stars and the [WR] planetary nebula we observed, none of our candidate sources have \ion{He}{2} emission lines, or are unusually hot for their luminosities, meaning it is unlikely we observed any RAWD.

\subsection{Expectations}\label{sec:expectations}

Our non-detection means that we can only set an upper limit to the number of RAWD in the SMC. To help establish these limits, we first describe a simple model and then turn to objects suggested to be RAWD.

\subsubsection {Comparison with Models}
To see what fraction of RAWD we should be able to detect, we created a simple model: a white dwarf with an extended envelope modeled as a black body and a main sequence star modeled with synthetic spectra from \citet{Pickles98}. For the companion, we try a range of masses 6.5--0.21$\,M_\odot$, corresponding to spectral types B5--M5) and set the binary separation such that they fill their Roche lobe. For the white dwarf, we assume either (1) that the envelope extends to the white dwarf's Roche lobe or (2) that it extends to half its Roche lobe. We also assume that the luminosity equals that of the upper edge of the steady burning region ($10^{4.64}\,L_\odot$; \citealt{Nomoto07}); the temperature then follows.

\begin{figure*}
\begin{center}$
\begin{array}{cc}
\includegraphics[width= .7 \linewidth]{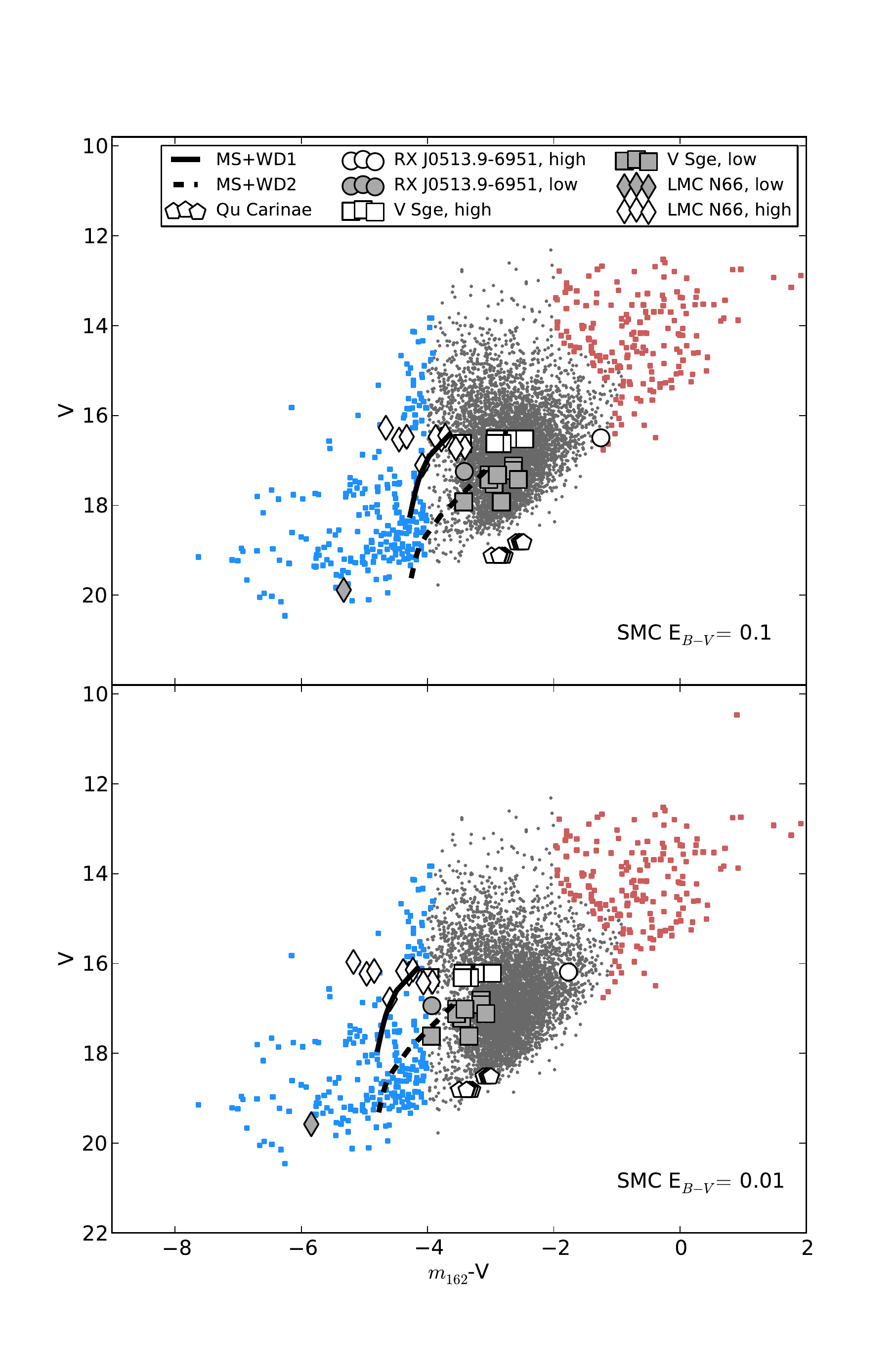}
\end{array}$
\end{center}
\caption{Same as Fig. \ref{cmd}, but also showing objects suggested to be rapidly accreting white dwarfs (as labeled). Two models are shown, with (1) the white dwarf filling its Roche lobe and (2) the white dwarf filling 50\% of its Roche lobe. The top panel assumes a conservatively high reddening, $E_{B-V}=0.1$, and the bottom panel assumes a very low reddening in the SMC, $E_{B-V}=0.01$.}
 \label{cmd_Vsge}
\end{figure*}

In order to compare our results to the observations, we assume a distance modulus of 18.9 to the SMC.  A source of uncertainty is the internal reddening of the SMC. Generally, for the SMC one has $E_{B-V}=0.02$ to 0.12, with about 0.02 to 0.04 coming from the Milky Way foreground and the rest from internal reddening \citep{Bessell91,Cornett97}.  While \citet{Cornett97} measure $E_{B-V}$ for objects that are in both the SMC UIT catalog and the AzV catalog of SMC OB stars \citep{Azzopardi82}, they assume a conservative $E_{B-V}=0.1$ for the SMC when discussing their sample as a whole. This value may be too large for RAWD, which should be substantially older than most OB stars and thus mostly outside of star forming regions, in areas of low extinction.  Hence, for our comparison in Figure~\ref{cmd_Vsge}, we show results both for the more conservative assumption, $E_{B-V}=0.1$ (top panel) and for a more optimistic one, $E_{B-V}=0.01$ (bottom panel).  For both cases, this is assumed to include Galactic reddening. We take the SMC extinction curve from \citet{Gordon03}.

From Figure~\ref{cmd_Vsge}, one sees that for the Roche-lobe filling model, the flux from the white dwarf dominates the system, with only a small companion contribution at the earliest spectral types.  For the half-filling case, however, the companion has a bigger influence, and the predictions merge with the main-sequence clump for companions earlier that  spectral type A0, corresponding to a mass of $\sim3.2\,M_\odot$.

\subsubsection{Comparison with Known RAWD Candidates}

There are no UIT observations of any of the known RAWD candidates --- the V Sge stars or LMC N66 (Section~\ref{sec:rawdcandidates}) --- but several of these objects do have spectra in the IUE archive. We can use these IUE observations to estimate UIT magnitudes fairly reliably, since IUE fluxes were used to calibrate UIT. UIT magnitudes are defined as: $m_{\lambda}=-2.5 \log F_{\lambda}-21.1$, where $F_{\lambda}$ is the flux in ${\rm\ erg \ cm^{-2}\ s^{-1}\AA^{-1}}$ and for the B5 filter $\lambda = 1615$\,\AA. For SMC objects, \citet{Parker98} find that the difference in magnitudes obtained from IUE spectra and UIT photometry is $0.04 \pm 0.25$ mag.

Combining these UV magnitudes with visual observations contemporary to the IUE observations, we can approximate where these objects would be found on Figure~\ref{cmd}. To transform the color and magnitude to the SMC, we first calculate absolute magnitudes with published distance and extinction estimates, and then estimate apparent magnitudes for the SMC. For the extinction curves, we use that of \citet{Gordon03} for the LMC and that of \citet{Seaton79} as fit by \citet{Fitzpatrick86} for the Milky Way.

\subsubsection{LMC N66}
While there are several IUE spectra of \object[LHA 120-N 66]{LMC N66} during quiescence, there is only one with a published, contemporary V magnitude observation from \citet{Pena95}. At this time, the object was both faint ($V=19.54$ at the distance of the SMC) and blue ($m_{162}-V=-5.89$). This suggests that if LMC N66 was able to pass the $U <19$ magnitude cut of our survey (which is likely given its blue SED), then it would have been selected as a candidate source.
 
LMC N66 was observed by the MACHO micro-lensing survey from Jan 1993 to Dec 1999, capturing the object's first outburst. A HST image of the system published in \citet{Pena04} shows a $\sim$ 5'' diameter nebula surrounding the central object LMC N66. There are two MACHO objects that are within 5'' of the central source, MACHO 61.9159.33 and MACHO 61.9159.42. Both sources show a rise and fall in their visual magnitude that coincides with the outburst. Comparing the positions given in the MACHO catalog (accurate to $\sim\!1$ arcsec) with the HST image, it seems that MACHO 61.9159.33 is most likely the central source, and MACHO 61.9159.42 is a knot in the surrounding nebula. MACHO 61.9159.42 is brighter in $V$ than MACHO 61.9159.33, but this is most likely due to strong emission from the nebular [\ion{O}{3}] 4959 and 5007 lines, which fall within the MACHO blue band and can contain up to ten percent of the total flux seen from a planetary nebula. 

LMC N66 is located at a distance modulus of 18.5 and $E_{B-V}=0.16\pm0.03$ \citep{Hamann03}. Using MACHO 61.9159.33, LMC N66 appears both brighter ($V=16.51\pm0.23$ at the distance of the SMC) and redder ($m_{162}-V=-3.63\mp0.40$) during outburst than during quiescence. This puts LMC N66 at the edge of the main sequence clump (see Figure~\ref{cmd_Vsge}).  Of the 9 observations, 4 fall within the region of our candidate sources using $E_{B-V}=0.1$ for the SMC and 8 of 9 using $E_{B-V}=0.01$.

 \subsubsection{V Sge}
\object[V* V Sge]{V Sge} has AAVSO observations dating back to 1910, and has been tracked continuously since 1930. This means that there are visual observations of the object within at least a few days of when every IUE spectra was taken. 

V Sge is reddened by $E_{B-V}=0.3\pm0.1$ \citep{Mader97}. While the distance to V Sge is poorly constrained, \citet{Hachisu03a} use models of the system to estimate it at $\sim\!3\,$kpc. During its optical hight state, V Sge has a similar visual luminosity to LMC N66 during outburst (corresponding to $V=16.46\pm0.053$ at the distance of the SMC). During its optical low state, it is brighter than LMC N66 during quiescence ($17.36\pm0.28$). Unlike LMC N66, the $m_{162}-V$ color of V Sge does not shift during its optical high and optical low state, remaining inside the central clump of Fig. \ref{cmd} and much redder than our candidate sources (see Fig. \ref{cmd_Vsge}).  It is not clear whether this is due to a large contribution from a cool disk or whether there is some internal extinction within the system. In either case, using $E_{B-V}=0.1$ for the SMC, none of the observations of V Sge fall into the region of our blue candidate sources. Using the optimistic $E_{B-V}=0.01$, 1 of 7 observations in its high state and 1 of 9 observations in its low state fall within this region. Thus, it is unlikely a star like V Sge would have been selected in our survey.

\subsubsection{RX J0513.9$-$6951}
\object[RX J0513.9$-$6951]{RX J0513.9-6951} has two IUE spectra, one during its optical high state, and one during its optical low state, that also have simultaneous MACHO observations. It is located at a distance modulus of 18.5 and $E_{B-V}=0.13\pm0.022$ \citep{Gansicke98}. RX J0513.9$-$6951 has a similar predicted V-band brightness to V Sge and LMC N66 during its optical high state, but has a much redder $m_{162}-V$ color. During its optical low state, its visual luminosity is similar to V Sge during its low state, but it has a bluer $m_{162}-V$ color, appearing at the edge of our main sequence clump (see Fig.~\ref{cmd_Vsge}).  Assuming $E_{B-V}=0.01$ for the SMC, it could be considered a candidate source, within uncertainties, but not with $E_{B-V}=0.1$.

\citet{Hachisu03a} claim that the differences in optical and X-ray variability between RX J0513.9$-$6951 and V Sge is due to differences in viewing geometry.  For V Sge, the inferred inclination is relatively close to edge-on, $i=70^{\circ}-80^{\circ}$, while RX J0513.9$-$6951 is suggested to viewed closer to pole-on, $i=20^{\circ}-30^{\circ}$.  With a lower inclination angle in RX J0513.9-6951, we observe less of the variability in the irradiated disk, and more of the change in photosphere of the WD. However, it is difficult to explain the color changes with this model, unless the expanding photosphere and winds of the white dwarf make the source appear very red for some reason.

\subsubsection{QU Car}
Several IUE spectra of \object[V* QU Car]{QU Car} are available, and we have visual observations within a few days of each thanks to it having been monitored continuously by AAVSO since 1987. A major uncertainty is the distance, which \citet{Drew03} put at ``$\sim\!2$\,kpc or more''. Using 2\,kpc and $E_{B-V}=0.1\pm0.015$, the object is much fainter and redder than V Sge and LMC N66. A distance of $\sim\!4\,$kpc is needed to bring its visual luminosity in agreement with V Sge and LMC N66.  Thus, either the published distance estimate is too low (which is plausible) or the system is not similar to V Sge.  In either case, given its relatively red color, it would not be considered a candidate source. 

\section{Limits to the Number of RAWD}
We did not detect any RAWD, and thus we are only able to establish an upper limit on the number of RAWD in the SMC, given our expectation of what such sources would look like, and the visibility limits of this survey. 

For any RAWD, the probability that we would find it can be written as three factors, $p_{\rm det}=p_{\rm infield} \times p_{\rm cand} \times p_{\rm fiber}$, where the three factors are the probabilities that the source is in one of the UIT fields, that it would be identified as a blue candidate source, and that it would be assigned a fiber, respectively.\footnote{Strictly, there is the additional probability of showing a sufficiently unusual spectrum.  We take this to be unity, given the known RAWD examples.}  

For a given number $n$ of RAWD sources, the probability of not detecting any of them is  $P(0|n)=(1 - p_{\rm det})^n$, and, conversely, the probability of detecting at least one is $1-P(0|n)$.  Thus the number of RAWD in the SMC for which at some confidence $1-P$ we should have detected one or more, is $n_{\rm max}=\log P/\log(1-p_{\rm det})$.  Below, we will quote 95\% confidence limits (i.e., set $P=0.05$).

Of the three probabilities that enter $p_{\rm det}$, two are straightforward to estimate.  The first is $p_{\rm infield}$, which is simply the fraction of the SMC covered by the UIT images, i.e., most of the central bar (see Fig.~\ref{circles}).  Since the UIT fields contain 30.6\% of all  $U<19$ entries in the MCPS catalog, and since the MCPS catalog covers the central $4.5 \times 4$ degrees of the SMC, which is the vast majority of the dwarf galaxy, we infer that $p_{\rm infield}=0.306$. The second is $p_{\rm fiber}$, which equals the fraction of 88.8\% of objects identified as candidate sources that were observed by our survey (due to constraints on fiber positioning and the fields of view of AAOmega and FLAMES).

The final piece is the probability $p_{\rm cand}$ that a RAWD would be identified as a candidate source. We can divide our predictions into three cases, based on different assumptions on what a RAWD looks like and how much reddening there is internal to the SMC.

In our first, most optimistic case, RAWD resemble faint Wolf-Rayet stars, with very blue $m_{162}-V$ colors, and thus always fall within the region of blue candidate sources in Fig.~\ref{cmd}.  Thus, for this case $p_{\rm cand}=1$ and $p_{\rm det}=0.27$, and our non-detection sets a limit of 10 RAWD in the SMC at the 95\% confidence level.

In the second case, RAWD resemble LMC N66.  If so, detectability depends on state and reddening.  LMC 66 should always be visible in quiescence, but in outburst the detectability depends on extinction: placed in the SMC with $E_{B-V}=0.1$, LMC N66 in outburst would be selected as a candidate source in 4 of 9 observations.  Hence, $p_{\rm cand}\simeq\frac{1}{2}+\frac{1}{2}\frac{4}{9}\simeq0.72$ and $p_{\rm det}=0.196$. This puts a limit of 14 RAWD similar to LMC N66 at the 95\% confidence level. Assuming $E_{B-V}=0.01$ for the SMC, it would be selected as a candidate source 8 of 9 times during outburst.  Hence, $p_{\rm cand}\simeq1$ and the limit is (nearly) the same as for our first case.

In the third, most pessimistic case, RAWD generally resemble V Sge.  If so, they would only be selected as candidates if the extinction is very low. V Sge spends about 180 days in its high state and 120 days in its low state. Assuming the most optimistic reddening of $E_{B-V}=0.01$, V Sge placed in the SMC would be selected in 1 of 7 observations in its high state and 1 of 9 observations in its low stare.  Hence, $p_{\rm cand}\simeq\frac{3}{5}\frac{1}{7}+\frac{2}{5}\frac{1}{9}\simeq0.13$, $p_{\rm det}=0.035$, and one infers a very loose upper limit of 83 RAWD. However, V~Sge would not be selected at all for our survey if $E_{B-V}$ were higher.

Thus, for the first two cases, we can put an upper limit of 10--14 RAWD in the SMC. However, the upper limits from the third case are not useful, since even under extremely favorable reddening conditions, they still are consistent with even the upper end of the range of 33--132 expected type Ia progenitors in the SMC (Section~\ref{sec:numbRAWD}).

\section{Conclusions}
We have described our attempt to find binaries containing RAWD among unusually UV-bright stars in the SMC.  We had hoped to find at least some, to empirically constrain the range of observed properties, but found none.  This either means that RAWD are substantially rarer than expected based on single-degenerate SN~Ia progenitor models, or that most RAWD are not very blue, e.g., because most are like the V Sge stars and few like LMC N66.  

To derive more stringent constraints would require using a more reliable indicator for a RAWD.  Apart from the unobservable extreme-UV flux, best may be the presence of high ionisation lines in the spectra, such as \ion{He}{2}, which are seen in all known examples (as well as in SSS).  Suitable candidates might be found using a narrow-band imaging survey.  Of course, \ion{He}{2} is seen in other objects as well, but most are interesting on their own accord (e.g., X-ray binaries, Wolf-Rayet stars) or also relevant for the SN~Ia progenitor question (e.g., symbiotics, progenitors on their own and a phase in the evolution to double degenerates). 

\acknowledgements
We acknowledge with thanks the variable star observations from the AAVSO International Database contributed by observers worldwide and used in this research.

This paper utilizes public domain data obtained by the MACHO Project, jointly funded by the US Department of Energy through the University of California, Lawrence Livermore National Laboratory under contract No. W-7405-Eng-48, by the National Science Foundation through the Center for Particle Astrophysics of the University of California under cooperative agreement AST-8809616, and by the Mount Stromlo and Siding Spring Observatory, part of the Australian National University. This paper also utilizes public domain data from the Ogle-II microlensing survey on-line photometric database.

Some of the data presented in this paper were obtained from the Mikulski Archive for Space Telescopes (MAST). STScI is operated by the Association of Universities for Research in Astronomy, Inc., under NASA contract NAS5-26555. Support for MAST for non-HST data is provided by the NASA Office of Space Science via grant NNX09AF08G and by other grants and contracts.

This research has made use of the VizieR catalogue access tool and the SIMBAD database, operated at CDS, Strasbourg, France. This research has also made use of NASA's Astrophysics Data System

{\it Facilities:} \facility{VLT:Kueyen (FLAMES)}, \facility{AAT (AAOmega), \facility{AAVSO}}

\bibliography{lib}

\appendix

\section{Appendix: Interesting objects found in the survey}\label{sec:interesting}
While our survey did not detect any sources that could be a rapidly accreting white dwarf, we did detect several objects that have similar properties, including two Wolf-Rayet stars and a [WR] planetary nebula. 




\paragraph{SMC USSC 131 = SMC WR11 } (Fig. 3.35). Observed with FLAMES and AAomega, we find NIII-V emission lines at $\sim460$ nm, a strong, wide \ion{He}{2} 4686 emission line, a weaker  H$\beta$ absorption line and a strong, wide H$\alpha$ emission line in the spectra of \object[2MASS J00520738-7235385 ]{SMC WR11}. H$\beta$ absorption lines are common in SMC WR stars --- all but SMC WR 4 exhibit early-type absorption spectra superposed n the canonical WN emission lines. While the presence of absorption lines can be signs of either a binary companion or absorption from thin stellar winds, SMC WR 11 shows no periodic radial velocity variations, which makes it likely that WR 11 is a single object \citep{Foellmi03a}. \citet{Foellmi03a} give a V magnitude of about 15.7, this is consistent with $V=15.845$ listed in MCPS. The absolute magnitude is $M_V\simeq-4.7$, using a distance modulus to the SMC of 18.9, which is similar to other SMC WN stars.  Given this and the spectrum, it is very unlikely that SMC 11 is a [WR] planetary nebula core.

\paragraph{SMC USSC 482 = SMC WR12} (Fig. 3.178) is the most recent Wolf-Rayet star to be discovered in the SMC, a (likely) single and hydrogen-rich WN3 star \citep{Massey03, Foellmi04}. Observed with AAOmega, we find strong, NIII-V emission lines at $\sim460$ nm, a strong, wide  \ion{He}{2} 4686 emission line, with a weak, asymmetric, H$\beta$ absorption and emission line and a strong, wide H$\alpha$ emission line in the spectra of \object[SMC AB 12]{SMC WR12}. As with other SMC WN stars, WR12 shows signs of a weak early-type absorption spectra.  \citet{Massey03} give $V=15.5$ for WR 12, somewhat brighter than the $V=16.158$ listed by MCPS.  This yields an absolute magnitude $M_V\simeq-4.0$, which is consistent with other SMC WN stars.  Given this and the spectrum, it is very unlikely that the object is a [WR] planetary nebula core.

\paragraph{SMC USSC 276 = LIN 302} is a [WC 8] Wolf-Rayet type planetary nebula core.  It is  similar to the galactic [WC 8] NGC 40, a very low-excitation, evolved nebula \citep{Barlow87, Pena97}. While it is likely that the central object of the planetary nebula is responsible for the UV emission seen with UIT, a bad match between the MCPS and UIT catalog gave a optical position which is offset from the central object, and thus our spectrum is dominated by the nebula. None the less, apart from the strong, narrow H$\beta$ and [OIII] 4959 and 5007 emission lines characteristic of a planetary nebula, we also detect weak, wide \ion{He}{2} 4686 and \ion{C}{3} 4647 emission lines from the [WC] core of \object[2MASS J00561947-7206586]{LIN 302}.

\end{document}